\documentclass[reprint]{revtex4-1}

\usepackage{amsmath}    
\usepackage{graphicx}   
\usepackage{verbatim}   
\usepackage{color}      
\usepackage{hyperref}   

\begin{document} 

\title{Valley relaxation of resident electrons and holes in a monolayer semiconductor: Dependence on carrier density and the role of substrate-induced disorder}

\author{Jing Li$^{1\#}$, M. Goryca$^{1\#}$, K. Yumigeta$^{2}$, H. Li$^{2}$, S. Tongay$^{2}$, S.~A. Crooker$^{1}$}
\affiliation{$^1$National High Magnetic Field Laboratory, Los Alamos National Laboratory, Los Alamos, New Mexico 87545, USA}
\affiliation{$^2$School for Engineering of Matter, Transport, and Energy, Arizona State University, Tempe, Arizona 85287, USA}

\date{\today}

\begin{abstract}
Using time-resolved optical Kerr rotation, we measure the low temperature valley dynamics of resident electrons and holes in exfoliated WSe$_2$ monolayers as a systematic function of carrier density.  In an effort to reconcile the many disparate timescales of carrier valley dynamics in monolayer semiconductors reported to date, we directly compare the doping-dependent valley relaxation in two electrostatically-gated WSe$_2$ monolayers having different dielectric environments. In a fully-encapsulated structure (hBN/WSe$_2$/hBN, where hBN is hexagonal boron nitride), valley relaxation is found to be monoexponential. The valley relaxation time $\tau_v$ is quite long ($\sim$10~$\mu$s) at low carrier densities, but decreases rapidly to less than 100~ns at high electron or hole densities $\gtrsim$2 $\times 10^{12}$~cm$^{-2}$.  In contrast, in a partially-encapsulated WSe$_2$ monolayer placed directly on silicon dioxide (hBN/WSe$_2$/SiO$_2$), carrier valley relaxation is multi-exponential at low carrier densities.  The difference is attributed to environmental disorder from the SiO$_2$ substrate. Unexpectedly, very small out-of-plane magnetic fields can increase $\tau_v$, especially in the hBN/WSe$_2$/SiO$_2$ structure, suggesting that localized states induced by disorder can play an important role in depolarizing spins and mediating the valley relaxation of resident carriers in monolayer transition metal-dichalcogenide semiconductors. 
\end{abstract}

\maketitle

Much of the current excitement surrounding the family of monolayer transition-metal dichalcogenide (TMD) semiconductors such as MoS$_2$ and WSe$_2$ derives from notions of utilizing ``valley pseudospin'' degrees of freedom as a basis for storing and processing quantum information \cite{Xu:2014, Schaibley:2016, Mak:2018}.  Although many of these ideas originated from long-standing considerations of valley degrees of freedom in other materials such as silicon, AlAs, and graphene \cite{Shkolnikov:2002, Takashina:2006, Rycerz:2007, Xiao:2007}, monolayer TMD semiconductors have revitalized interests in ``valleytronics'' due to the ease with which specific valleys in momentum space (namely, the $K$ and $K'$ points in the Brillouin zone) can be selectively populated and probed simply by using/detecting right- or left-circularly polarized light.  This convenient set of valley-specific optical selection rules in the monolayer TMDs  --which do not exist in most conventional semiconductors or in graphene-- arises from their lack of inversion symmetry, and large spin-orbit coupling \cite{Yao:2008, Xiao:2012}. 

Analogous to the keen interest in electron, hole, and exciton spin relaxation during the early days of semiconductor spintronics \cite{OO, SS, Dyakonovbook}, measurements of valley relaxation in monolayer TMDs are currently a focus of attention for potential applications in valleytronics.  Considerable initial attention followed from early photoluminescence (PL) studies in 2012 demonstrating that when certain monolayer TMDs were photoexcited by circularly-polarized light, the resulting PL was strongly co-circularly polarized \cite{Sallen:2012, Zeng:2012, Mak:2012, Cao:2012}, suggesting that valley degrees of freedom might be especially robust and long-lived.  However, because PL necessarily originates from the optical recombination of transient excitons, co-polarized PL is equally consistent with very short exciton lifetimes (\textit{i.e.}, even shorter than the timescale of fast exciton valley scattering).  And indeed, very fast exciton and trion PL lifetimes in the 1-100~ps range were subsequently revealed in time-resolved PL studies \cite{Wang:2014, Robert:2016, Wang:2018}. 

While the dynamics of excitons and exciton valley scattering in monolayer TMDs \cite{Wang:2018, Mai:2014, Zhu:2014, Singh:2016, Hao:2016, Yu:2014,  DalConte:2015, Yan:2017, Plechinger:2016, Molina:2017} are surely very important for certain opto-electronic applications, a different and arguably more relevant question for many putative valleytronic applications is: ``What are the intrinsic valley relaxation timescales of the \textit{resident} electrons and holes that exist in $n$-type doped and $p$-type doped TMD monolayers?''.  In this case, electron-hole exchange interactions -- which can be a very efficient valley scattering and decoherence mechanism for polarized excitons \cite{Wang:2018, Mai:2014, Zhu:2014, Singh:2016, Hao:2016, Yu:2014,  DalConte:2015, Yan:2017, Plechinger:2016, Molina:2017} -- are absent.  Therefore, much longer valley relaxation timescales may be expected for resident electrons and holes, because the spin-orbit splitting and resulting spin-valley locking in the conduction bands and in (especially) the valence bands mandates that valley scattering at low temperatures requires not only a large momentum change ($K \leftrightarrow K'$), but \textit{also} a spin flip \cite{Xiao:2012}. 

Similar to the way in which long spin lifetimes of resident electrons and holes  enabled spin-based device demonstrations in conventional semiconductors and graphene \cite{Lou:2007, Kroutvar:2004, Appelbaum:2007, Han:2014}, it is therefore the valley relaxation of the resident electrons and holes in doped TMD monolayers and heterostructures that will likely determine the functionality of many valley-based device concepts. To this end, systematic studies of resident carrier valley dynamics in doped TMD monolayers are essential, particularly as a function of carrier density, temperature, and overall material quality. 

Experimental studies along these lines commenced in about 2015, typically using polarization-resolved optical pump-probe methods such as time-resolved circular dichroism and Kerr/Faraday rotation -- techniques that had been developed and refined in the preceding decades to explore spin lifetimes and spin coherence in conventional III-V and II-VI semiconductors \cite{Crooker:1997, Greilich:2006}.  Initial measurements focused on MoS$_2$, WS$_2$, and WSe$_2$ monolayers grown by chemical vapor deposition (CVD), which were unintentionally doped with either electrons or holes \cite{Yang:2015, Hsu:2015, Song:2016, McCormick:2017}. These studies typically revealed multi-exponential valley relaxation with longer timescales on the order of several nanoseconds at low temperatures (\textit{i.e.}, roughly $10 - 100 \times$ longer than the measured recombination times of the ``bright'' excitons and trions \cite{Wang:2014, Robert:2016}, and also the nominally-forbidden ``dark'' excitons and trions \cite{Robert:2017, Zhang:2017, Liu:2019, Molas:2019, Chen:2020}). While very encouraging, the CVD-grown monolayers studied in \cite{Yang:2015, Hsu:2015, Song:2016, McCormick:2017} were not charge-adjustable, and moreover they exhibited broad optical absorption lines (many tens of meV) indicating significant inhomogeneous broadening, likely due to the fact that they were grown by CVD, directly on SiO$_2$ or sapphire substrates, with surfaces that were not encapsulated or passivated.  

\begin{figure*}[t]
\centering
\includegraphics[width=1.6\columnwidth]{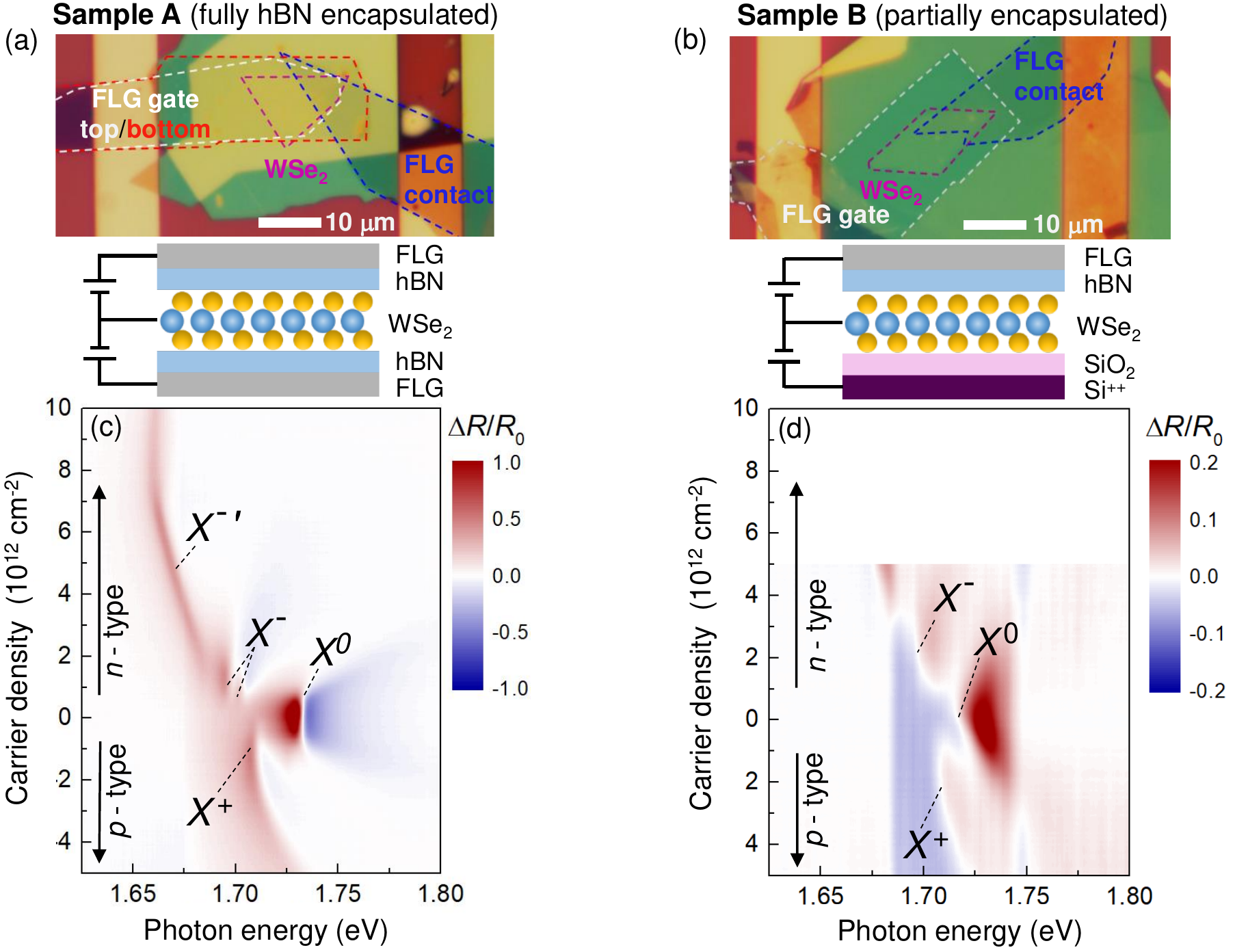}
\caption{\label{Fig1} (a) Optical micrograph and cross-sectional schematic of ``sample A'', which is a \textit{fully}-encapsulated WSe$_2$ monolayer (hBN/WSe$_2$/hBN). All hBN layers are $\sim$30~nm thick. The resident carrier density is electrostatically tuned via few-layer graphite (FLG) flakes, which serve as the top/bottom gates and contact electrode. The top/bottom gates are held at the same potential and swept together. (b) An optical micrograph and cross-sectional schematic of ``sample B'', which is a \textit{partially}-encapsulated WSe$_2$ monolayer placed directly on a 300~nm thick silicon dioxide layer (hBN/WSe$_2$/SiO$_2$). Here, the bottom Si$^{++}$ gate is held fixed at $-15$~V, and the carrier density is varied by the top FLG gate. (c) and (d) show intensity maps of the carrier-density-dependent reflectance spectra $\Delta R/R_0$ in Sample A and B, respectively. Electron- and hole-doped regimes are clearly visible in both structures, separated by a nominally charge-neutral regime in which the neutral exciton $X^0$ resonance is visible. Negatively-charged and positively-charged exciton features (\textit{i.e.}, $X^-$ and $X^+$ trions) appeared in the electron-doped ($n$-type) and hole-doped ($p$-type) regimes, respectively. Sample A exhibits considerably higher optical quality than sample B, as evidenced by its much narrower $X^0$ resonance (6~meV vs. 35 meV width) and trion resonances, and by the smaller gating range over which the charge-neutral regime appears. Carrier densities were calibrated by the known thicknesses and dielectric constants of the hBN and SiO$_2$ layers, and also (in sample A) by the ``kink'' in the $X^{-\prime}$ resonance at high electron density \cite{Wang:2017}. All measurements were performed at 5.8 K.}
\end{figure*}

Subsequent studies then focused on carrier valley relaxation in mechanically-exfoliated (but still unpassivated) WSe$_2$ monolayers that exhibited improved optical quality, and which afforded some degree of charge tunability via electrostatic gating \cite{Dey:2017}.  Here, $\sim$100~ns valley relaxation of resident electrons was found in the heavily $n$-type doped regime, while a very long microsecond-duration valley relaxation of holes was observed when the same monolayer was gated $p$-type (although a systematic dependence on carrier density was not performed). Similarly, pump-probe measurements of WSe$_2$/MoS$_2$ \textit{bi}-layers, in which photoexcited holes quickly migrated to the WSe$_2$ layer and then relaxed independently from the electrons, also showed long microsecond valley relaxation of holes \cite{Kim:2017}. 

In parallel, at about this time (2017) it became widely appreciated that the optical quality of TMD monolayers could be significantly improved via full encapsulation between slabs of exfoliated hexagonal boron nitride (hBN) \cite{Cadiz:2017, Ajayi:2017, Wang:2017}. This key advance made it possible to achieve TMD monolayers with not only exceptionally narrow (a few meV) exciton linewidths approaching the homogeneous limit, but also with the ability to be continuously gated all the way from the heavily electron-doped regime to the heavily hole-doped regime \cite{Wang:2017, VanTuan:2019}, likely owing to a reduction of defects, traps, and localized states in the WSe$_2$ forbidden band gap.  Using hBN-encapsulated structures, \textit{mono-}exponential valley relaxation of resident electrons and holes was observed, with timescales of order 100~ns (for electrons) to several microseconds (for holes) \cite{Jin:2018, Goryca:2019}. Crucially, these long timescales were validated by passive optical detection of the thermodynamic valley fluctuations (`valley noise') of the resident carriers in thermal equilibrium (\textit{i.e.}, no pump laser or excitation was used) \cite{Goryca:2019}, confirming that the long decays originate from the intrinsic valley relaxation of the resident electrons and holes themselves, and not from, \textit{e.g.}, any type of long-lived dark or trapped exciton \cite{Volmer:2017}. 

As outlined above, a wide range of electron and hole valley relaxation timescales have been reported to date, which is due in part to the significant improvement in the quality and passivation of TMD monolayers over the past several years, and also because different studies reported on samples with different specific carrier densities. Here, we attempt to provide a more unified understanding of resident carrier valley relaxation in monolayer TMD semiconductors, by directly comparing the detailed doping dependence of valley relaxation in two electrostatically-gated WSe$_2$ monolayers having different surrounding dielectric environments. In a \textit{fully-encapsulated} WSe$_2$ monolayer (hBN/WSe$_2$/hBN), valley relaxation is found to be almost perfectly monoexponential. The valley relaxation time $\tau_v$ is quite long ($\sim$10~$\mu$s) at low carrier densities, but decreases rapidly to less than 100~ns at high electron or high hole densities exceeding $\sim$2 $\times 10^{12}$~cm$^{-2}$.  In contrast, in a \textit{partially-encapsulated} WSe$_2$ monolayer placed directly on a silicon dioxide substrate (hBN/WSe$_2$/SiO$_2$) -- another common assembly motif -- carrier valley relaxation is found to be multi-exponential at low densities, exhibiting a faster initial decay followed by a slow ($\sim$10~$\mu$s) relaxation. The difference is attributed to in-gap states in the WSe$_2$ introduced by dielectric disorder in the SiO$_2$ substrate, which has an especially strong influence on monolayer TMDs in the low carrier density regime. Similar to the fully-encapsulated case, however, this slow relaxation accelerates by over two orders of magnitude with increasing electron or hole doping.  Increasing temperatures also accelerate valley relaxation, and  -- crucially -- the unexpected influence of small perpendicular magnetic fields is measured and discussed in the context of localized in-gap states that can depolarize spins and thereby mediate carrier valley relaxation. 

Figure 1 shows the two different electrostatically-gated WSe$_2$ monolayer structures measured in this work. To ensure accurate  comparisons, both structures incorporate a single WSe$_2$ monolayer that was mechanically exfoliated from the same high-quality bulk crystal grown by the self-flux method, in which the density of lattice defects is much lower than in bulk crystals grown by chemical vapor transport \cite{Edelberg:2019}. Both structures were assembled using the same van der Waals dry stacking method to provide clean interfaces between the WSe$_2$ and its adjacent dielectric materials. 

``Sample A'' (Fig. 1a) is representative of modern structures exhibiting very high optical quality: it is a \textit{fully} hBN-encapsulated WSe$_2$ monolayer (hBN/WSe$_2$/hBN), with top/bottom gates and a contact electrode made from flakes of few-layer graphite (FLG). Here, the complete encapsulation by pristine (not post-processed) exfoliated hBN flakes provides an atomically-smooth and largely defect-free dielectric environment \cite{Dean:2010} entirely surrounding the WSe$_2$ monolayer, which leads to very narrow features in optical spectra, as discussed below.  Full hBN encapsulation is now widely recognized to yield monolayer TMD structures with excellent optical properties and a minimum of inhomogeneous broadening and disorder \cite{Cadiz:2017, Ajayi:2017, Raja:2019, Rhodes:2019, Martin:2020}.

In contrast, ``Sample B'' (Fig. 1b) is representative of a different, but nonetheless very common, type of van der Waals structure: it is a \textit{partially}-encapsulated WSe$_2$ monolayer that is placed directly on silicon dioxide (hBN/WSe$_2$/SiO$_2$). Here,  gating is achieved via a top FLG gate electrode and the underlying heavily-doped Si substrate.  In this case, the WSe$_2$ monolayer is in direct contact with the SiO$_2$, and is therefore in intimate proximity to surface defects, dangling bonds, dielectric disorder, and surface roughness, with a consequent reduction in the optical quality of the TMD monolayer.  

\begin{figure*}[t]
\centering
\includegraphics[width=1.8\columnwidth]{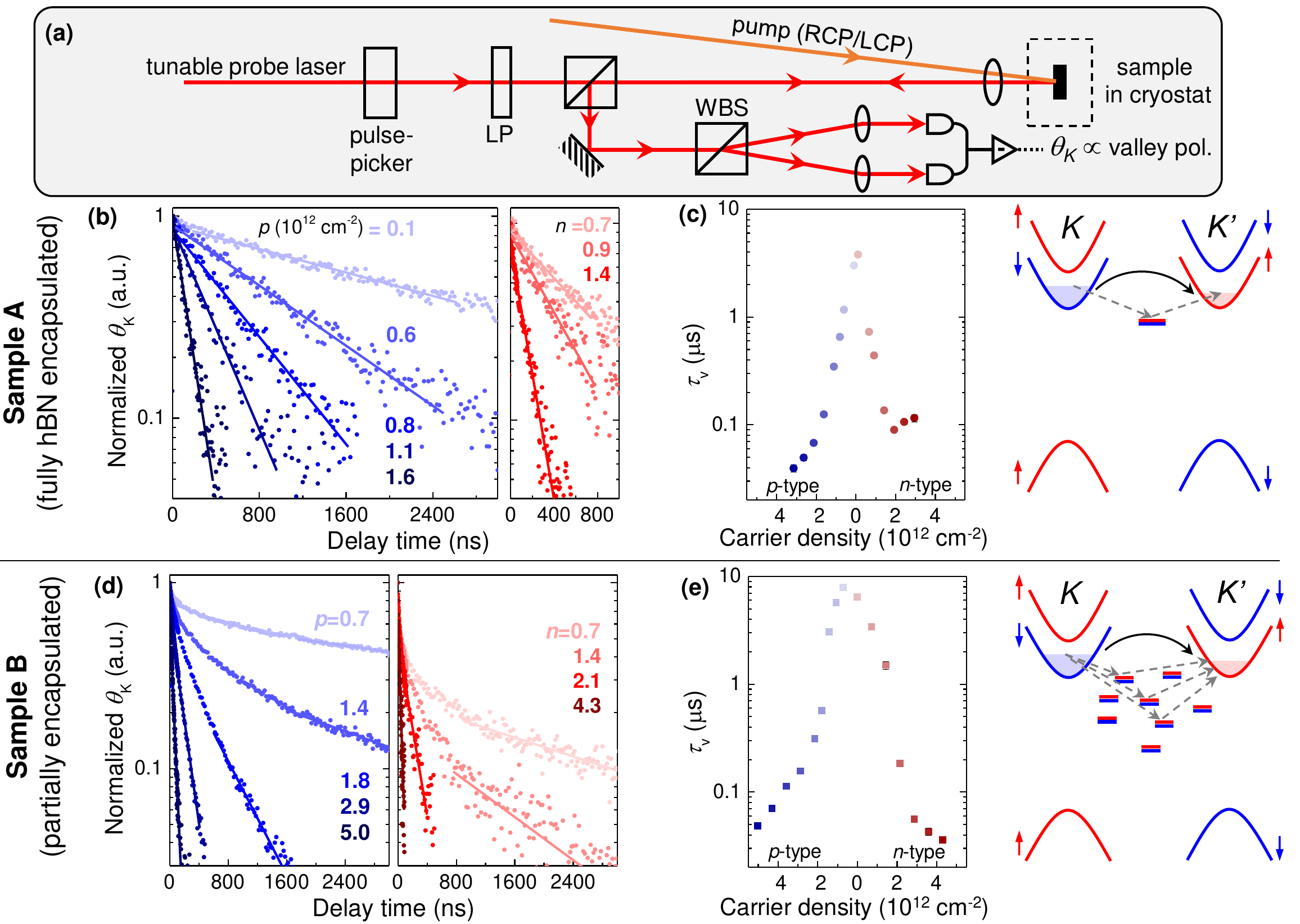}
\caption{\label{Fig2} (a) Schematic of the time-resolved Kerr rotation (TRKR) experiment.  A non-equilibrium spin-valley polarization of the resident carriers in the WSe$_2$ monolayer is induced by circularly-polarized (RCP/LCP) pump pulses from a 645~nm (1.922~eV) diode laser. The induced polarization is detected by the Kerr rotation $\theta_K$ imparted on time-delayed and linearly-polarized probe pulses from a wavelength-tunable Ti:sapphire laser that was typically tuned to the peak of the induced Kerr response (\textit{cf.} Fig. 3). A pulse-picker (PP) reduced the repetition rate of the Ti:sapphire laser to allow accurate measurement of long valley decays, and the pump laser is synchronized to the PP \cite{Dey:2017}. LP: linear polarizer, WBS: Wollaston beam splitter. (b) TRKR signals from  sample A (fully-encapsulated), at different electron ($n$) and hole ($p$) carrier densities. The decays are monoexponential, except for a very slight deviation at short timescales at the lowest densities. The signals reveal the relaxation of optically-induced valley polarization of resident carriers, which in monolayer TMDs is intimately linked to carrier spin polarization due to spin-valley locking \cite{Xiao:2012}. (c) The corresponding valley relaxation time $\tau_v$ versus carrier density; $\tau_v$ is sharply peaked at low carrier density. (d) TRKR from  sample B (partially-encapsulated). Here, at low carrier densities the valley relaxation exhibits a faster initial decay that is followed by a slower exponential decay. (e) The lifetime $\tau_v$ of the slow valley relaxation is also peaked at low carrier density, similar to sample A. All measurements were performed at 5.8~K. The diagrams depict localized in-gap states (with nearly degenerate spin levels) that can potentially mediate the valley depolarization of resident carriers, especially at small carrier densities.}
\end{figure*}

The difference in optical and material properties between the two structures is readily apparent in the gate-dependent maps of optical reflectivity, shown in Figs. 1(c,d). These maps show normalized reflection spectra ($\Delta R / R_0$) acquired at low temperature (5.8~K), as the resident carrier density in the WSe$_2$ monolayers is tuned from heavily electron doped ($n$-type), to nominally undoped (charge neutral), to heavily hole doped ($p$-type). For clarity, throughout this paper we use ``$n$'' and ``$p$'' when referring to electron and hole densities, respectively. High electron densities ($n$) and high hole densities ($p$) exceeding $5 \times 10^{12}$~cm$^{-2}$ are achieved in both structures, thanks to the dual-gating configuration. Importantly, no hysteresis as a function of gate voltage or slow drifts were observed in the measured optical properties of the WSe$_2$ \cite{JingLi:2020}, likely due to the use of pristine (never post-processed) hBN encapsulation layers, and because (in sample B) the bottom Si$^{++}$ gate voltage is never varied. 

In the fully-encapsulated sample A (Fig. 1c), a series of sharp exciton resonances are observed as a function of carrier density. The neutral exciton resonance ($X^0$) at $\sim$1.733~eV appears only in a narrow gating region around the charge neutrality point. When $n$ and $p$ exceed $\sim$0.3 $\times 10^{12}$~cm$^{-2}$, strong negatively-charged exciton ($X^-$) and positively-charged exciton ($X^+$) resonances appear, respectively, at $\sim$1.700~eV and 1.710~eV. The appearance of charged excitons (also called `trions' or `attractive polarons' in the literature) in optical reflectivity spectra indicates a substantive change in the available joint density of states of the WSe$_2$ monolayer, and therefore the absorption oscillator strength, that is due to the occupation of the conduction bands (CB) and valence bands (VB) by resident carriers. At large electron doping ($n > 2 \times 10^{12}$~cm$^{-2}$), an additional resonance appears at lower energy, which further redshifts with increasing $n$. Although the nature of this resonance is not yet fully understood \cite{VanTuan:2017, Wang:2017}, in keeping with recent studies we label this peak as $X^{-\prime}$. Importantly, these reflectivity spectra are in excellent agreement with other recent measurements of gate-dependent absorption and reflection of hBN-encapsulated WSe$_2$ monolayers \cite{Wang:2017, VanTuan:2019}.

In the partially-encapsulated sample B (Fig. 1d), we observe a qualitatively similar dependence of $\Delta R/R_0$ on carrier density. However, the exciton resonances are spectrally much broader and weaker (note the different intensity scale). The neutral exciton $X^0$ is clearly visible near the charge neutrality point, but is spectrally broad ($\sim$35~meV, versus $\sim$6~meV in the sample A). $X^\pm$ charged exciton features are also broader, and do not appear until the carrier density exceeds $\sim$1 $\times 10^{12}$~cm$^{-2}$, suggesting the presence of ample states in the nominally forbidden band gap of the WSe$_2$ monolayer, likely introduced by the close proximity to the SiO$_2$ substrate.

Figure 2(a) depicts the time-resolved Kerr rotation (TRKR) experiment that was used to measure the valley dynamics of the resident carriers. The gated WSe$_2$ structures were mounted on the cold finger of a small optical cryostat. A pulsed diode laser generated short (80~ps) optical pump pulses at 645~nm (1.922~eV), which were then circularly polarized and focused to a 10 $\mu$m diameter spot on the sample. Owing to the valley-specific optical selection rules in monolayer WSe$_2$, the above-gap pump photoexcites valley- and spin-polarized electrons and holes into the monolayer, which rapidly relax to the band edges and form bright and dark excitons and trions, which subsequently scatter and recombine on short ($\lesssim$1~ns) timescales \cite{Wang:2014, Robert:2016, Robert:2017, Zhang:2017, Liu:2019, Molas:2019, Chen:2020, Wang:2018}.  Invariably, some of the possible recombination pathways involved phonon-assisted transitions, nonradiative recombination, and the existing resident carriers -- the net effect of which induces a nonequilibrium spin/valley polarization to the Fermi sea of resident carriers. 

The subsequent relaxation of this pump-induced valley polarization on timescales longer than a few hundred picoseconds was detected via the optical Kerr rotation $\theta_K (t)$ imparted on time-delayed and linearly-polarized probe pulses (150~fs) from a wavelength-tunable Ti:sapphire laser focused to a small 3~$\mu$m diameter spot on the sample.  To enable the accurate and unambiguous measurement of the long (0.1-10 $\mu$s) valley relaxation observed in this work, it was necessary to reduce the repetition rate of the Ti:sapphire laser by an acousto-optic pulse-picker, so that the time interval between excitation cycles was longer than the measured relaxation time.  The time delay between the pump and probe pulses was synchronized and controlled by an electronic timing delay \cite{Dey:2017}. The pump pulses were modulated between right- and left-circular polarization by a photoelastic modulator, and the pump-induced Kerr rotation was detected by balanced photodiodes and measured via standard lock-in techniques. The probe laser was typically tuned in photon energy to the peak of the induced Kerr response, which coincided with the spectral position of the $X^\pm$ charged exciton resonances, as discussed in more detail below.

\begin{figure}[b]
\centering
\includegraphics[width=0.98\columnwidth]{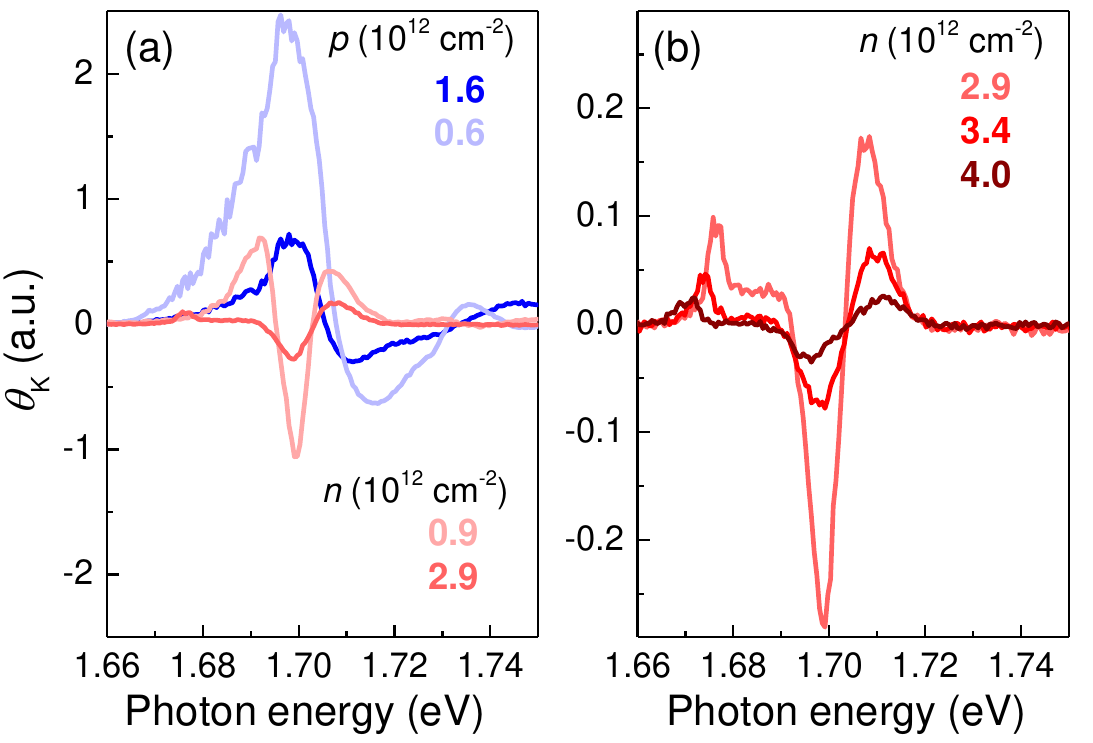}
\caption{\label{Fig3} Spectra showing the optically-induced Kerr rotation $\theta_K$ versus probe laser energy, at selected electron and hole densities in sample A. Here, both the pump and probe lasers are operated in continuous wave (cw, not pulsed) mode. The circularly-polarized cw pump laser (1.922~eV) generates a small steady-state nonequilibrium spin-valley polarization of the resident carriers.  The influence of this carrier polarization is detected as a function of photon energy by the wavelength-tunable narrowband cw probe laser. The dispersive lineshapes of the $\theta_K$ resonances are centered around 1.71 eV, which corresponds to the charged exciton resonances (see Fig. 1c). The additional resonance appearing at lower energy ($\sim$1.675~eV) in the heavily electron-doped regime corresponds to the $X^{-\prime}$ feature. We emphasize that Kerr rotation derives from differences between right- and left-circularly polarized optical absorption and reflection, and as such is a direct probe of changes in real oscillator strength.}
\end{figure}

The carrier-density-dependent valley relaxation in both samples is shown in Figs. 2(b-e). Red-shaded and blue-shaded traces show measured decays of $\theta_K (t)$ in the electron-doped ($n$-type) and hole-doped ($p$-type) regimes, respectively. In the fully-encapsulated sample A, $\theta_K$ was observed to decay almost perfectly mono-exponentially at all carrier densities (Fig. 2b), suggesting that a single relaxation mechanism dominates.  We note that monoexponential decays were also observed in recent time-resolved studies of carrier valley dynamics in high-quality (fully hBN encapsulated) structures \cite{Goryca:2019, Jin:2018}. The valley lifetime $\tau_v$ that we observe in sample A is very long at low carrier density (several microseconds), in agreement with recent time-resolved measurements \cite{Dey:2017, Goryca:2019, Jin:2018, Kim:2017}. However, we find that $\tau_v$ drops precipitously with increasing electron or hole doping, to values on the order of 100~ns at high electron or high hole densities $\geq 2 \times 10^{12}$~cm$^{-2}$. We emphasize that even this comparatively rapid decay (100~ns) is still approximately 2-3 orders of magnitude longer than the measured recombination time scales of dark excitons and trions (0.1-1~ns) in monolayer WSe$_2$ \cite{Robert:2017, Zhang:2017, Liu:2019, Molas:2019, Chen:2020}, strongly supporting its interpretation as the slow valley relaxation of resident carriers. The complete set of measured valley lifetimes in sample A as a function of resident carrier density are summarized in Fig. 2c. As just described, $\tau_v$ exhibits a sharp maximum at low carrier density.

\begin{figure*}[t]
\centering
\includegraphics[width=1.8\columnwidth]{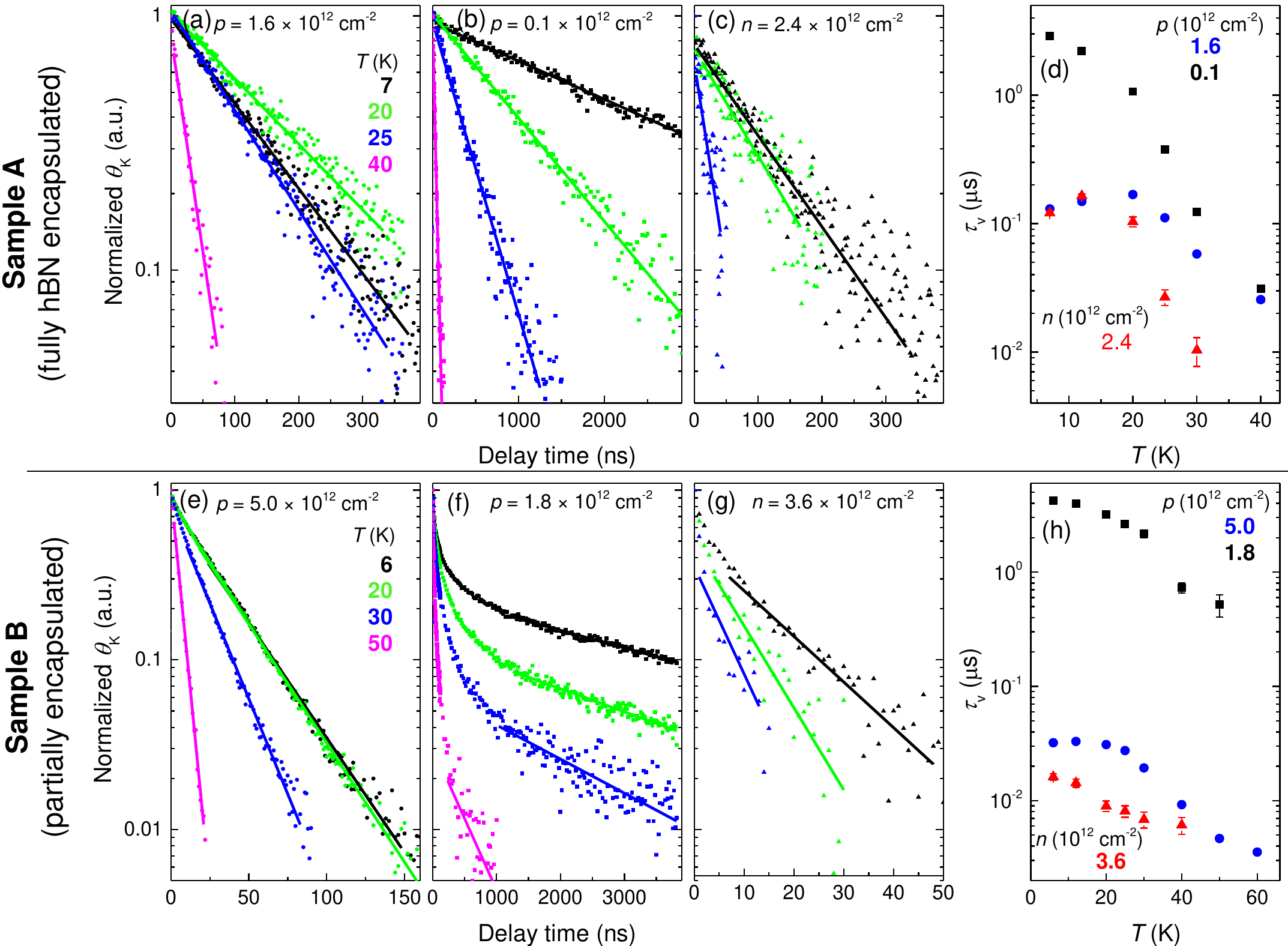}
\caption{\label{Fig4}(a-c) Measured valley relaxation (normalized) in sample A, at different temperatures and at three different carrier densities (\textit{i.e.}, in the hole-doped, approximately charge-neutral, and electron-doped regime). Points show  data; lines show exponential fits. (d) $\tau_v$ versus temperature in sample A, for different carrier densities. (e-h) Same, but for sample B.}
\end{figure*}

In contrast, valley dynamics in the partially-encapsulated sample B (Fig. 2c) are found to be multi-exponential, especially at lower carrier densities below $\lesssim$2 $\times 10^{12}$~cm$^{-2}$, suggesting multiple relaxation mechanisms at play. In this low-density regime, a faster initial relaxation is typically followed by a slower and nearly mono-exponential decay.  These multi-exponential decays are similar to time-resolved measurements reported in early studies of CVD-grown and/or unencapsulated TMD monolayers having broader exciton linewidths and reduced optical quality \cite{Yang:2015, Hsu:2015, Song:2016, McCormick:2017}. We associate the slower decay at long time delays with the intrinsic valley relaxation time $\tau_v$, and again find long valley depolarization on the order of 10~$\mu$s at low carrier densities (\textit{i.e.}, similar to sample A).  However, $\tau_v$ accelerates by over two orders of magnitude with increasing $n$- and $p$-type doping, again similar to sample A (moreover, we note that valley relaxation becomes more closely mono-exponential).  The dependence of $\tau_v$ on carrier density in sample B (Fig. 2e) roughly follows that observed in sample A (Fig. 2c), although its decrease with $n$ and $p$ is less abrupt. 

We emphasize that samples A and B were fabricated using identical WSe$_2$, hBN, and graphite source materials, the same mechanical exfoliation protocol, and the same van der Waals dry-stacking setup and process. Therefore we associate the qualitatively different carrier valley dynamics observed in sample B (that is, multi-exponential decay) to an extrinsic origin, likely due to dielectric disorder and inhomogeneity introduced by the underlying SiO$_2$.  Because SiO$_2$ is a convenient and common substrate for a variety of 2D material platforms, these disorder potentials and related deleterious effects have been studied carefully over the past years, particularly in relation to the properties of SiO$_2$-supported graphene \cite{Zhang:2009, Dean:2010, Decker:2011}, and more recently in relation to the optical and electronic properties of TMD monolayers \cite{Shin:2016, Borys:2017, Raja:2019, Rhodes:2019}.  For example, scanning tunneling microscopy studies \cite{Shin:2016} revealed charge puddles in TMD monolayers on SiO$_2$, wherein the energies of the conduction band minima and valence band maxima varied by hundreds of meV over lateral length scales of order 10~nm.  In comparison, pristine hBN provides a largely defect- and inhomogeneity-free platform for both graphene and TMD monolayers \cite{Dean:2010, Decker:2011, Rhodes:2019}. As we discuss in more detail below, our time-resolved measurements of carrier valley depolarization in monolayer WSe$_2$ are consistent with a larger degree of inhomogeneity in the SiO$_2$-supported sample B, and the formation of localized in-gap states arising from this environmental disorder, which in turn can play a role in depolarizing spins and mediating valley relaxation. 

\begin{figure*}[t]
\centering
\includegraphics[width=1.8\columnwidth]{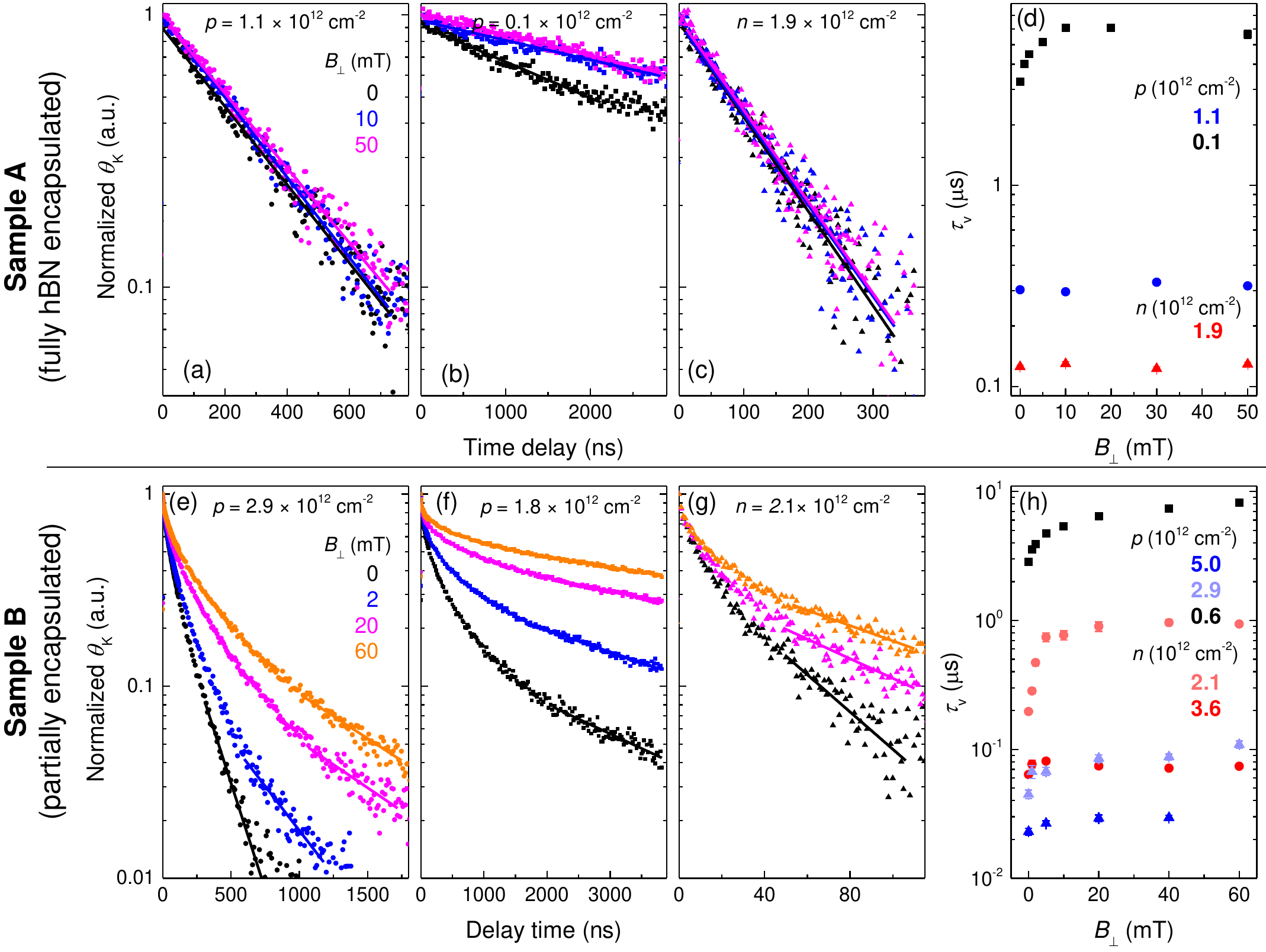}
\caption{\label{Fig5} (a-c) Measured valley relaxation (normalized) in sample A, at different applied out-of-plane magnetic fields $B_\perp$, at three different carrier densities (in the hole-doped, approximately charge-neutral, and electron-doped regimes). All data were taken at low temperature (5.8~K).  Points show data; lines show exponential fits. (d) $\tau_v$ versus $B_\perp$ in sample A, for different carrier densities. (e-h) Same, but for sample B, where $B_\perp$ has a more marked influence on $\tau_v$ at all carrier densities.}
\end{figure*}

Before turning to the role of substrate-induced disorder, we discuss the spectral dependence of the measured Kerr rotation signals.  As noted above, all the TRKR measurements were performed using above-gap pump pulses with fixed wavelength (645~nm, or 1.922~eV), while the probe pulses were tuned in wavelength near the peak of the induced Kerr-rotation response, which occurs in the spectral vicinity of the $X^-$ and $X^+$ charged exciton resonances.  For completeness, the full spectral dependence of the induced Kerr response from sample A is shown in Fig. 3, at different representative electron and hole densities. Here, the experiment is essentially the same as for TRKR (see Fig. 2a), except that the pump diode laser and the tunable probe laser are now operated in continuous-wave (cw) mode.  Thus, the measured $\theta_K$ is the induced Kerr rotation in response to a small \textit{steady-state} (but still non-equilibrium) valley polarization of the resident carriers comprising the Fermi sea.  

In our experience, a comprehensive understanding of the absorption spectrum and the optically-induced Kerr spectrum is essential in studies of this type.  This is because, as can be seen from Fig. 3, relatively small variations of probe laser's photon energy with respect to the system's underlying absorption resonances can result in Kerr signals having significantly different magnitude and even opposite sign (although measured timescales are typically not affected, provided the signals derive from the same absorption resonance).  Changes in signal magnitude and sign may arise trivially, due for example to exciton resonances that shift in energy (with respect to a fixed probe laser) as a function of temperature or carrier density or strain.  The spectra in Fig. 3 show that the optically-induced Kerr response exhibits a clear dispersive antisymmetric resonance centered near $\sim$1.71~eV, which coincides with the spectral position of the $X^-$ and $X^+$ charged exciton resonances (\textit{cf.} Fig. 1).  As discussed and shown previously \cite{Dey:2017, Goryca:2019}, this behavior is in line with expectations: a non-zero valley polarization of resident carriers is anticipated to manifest most prominently in the circular polarization of the bright $X^\pm$ charged exciton resonances, since the very existence of charged excitons depends explicitly of the presence of resident carriers in the monolayer. 

Figure 4 shows how the measured carrier valley relaxation changes with temperature (up to $\sim$50~K), at various densities, in both samples A and B. Overall, $\tau_v$ decreases as temperature increases, and qualitatively similar temperature-dependent trends were also observed in previous studies of carrier valley relaxation \cite{Yang:2015, Song:2016, Kim:2017, Goryca:2019, Ersfeld:2020}. We note, however, that while $\tau_v$ varies significantly with temperature at very low doping (in sample A), it is relatively insensitive to temperatures $\lesssim$20~K at higher carrier densities (\textit{e.g.}, $p= 1.6 \times 10^{12}$~cm$^{-2}$, or $n= 2.4 \times 10^{12}$~cm$^{-2}$, as shown in Figs. 4a and 4c.).  In contrast, $\tau_v$ in sample B exhibits a marked temperature dependence over a wider range of carrier densities. In general, temperature-dependent valley relaxation of resident carriers can be expected if carrier-phonon interactions promote spin-flip intervalley scattering \cite{Molina:2017, Ochoa:2013, Song:2013, Gunst:2016}, either directly or via Elliot-Yafet and Dyakonov-Perel spin relaxation mechanisms \cite{WangWu:2014, Ochoa:2014}, which have recently been reevaluated using \textit{ab initio} methods for 2D materials lacking spatial inversion symmetry \cite{Xu:2020}.

Small applied magnetic fields provide important additional insight. Figure 5 shows $\tau_v$ measured in both samples A and B, at low temperature (5.8~K), as a function of small applied out-of-plane magnetic field $B_\perp$ up to $\sim$50~mT.  In sample A, $\tau_v$ is insensitive to field except at  very small carrier densities (\textit{e.g.}, as shown in Fig. 5b where $p = 0.1 \times 10^{12}$~cm$^{-2}$), where $\tau_v$ approximately doubles, but then saturates by 10~mT. For larger electron or hole doping, $\tau_v$ is insensitive to $B_\perp$ (see Figs. 5a and 5c). In marked contrast, in sample B both $\tau_v$ and the overall valley decays show a considerable variation with $B_\perp$ at all measured electron and hole carrier densities -- and in particular, the faster component is rapidly suppressed with $B_\perp$ [see Figs. 5(e-g)].

The sensitivity of $\tau_v$ to small $B_\perp$ is perhaps somewhat surprising in view of the large spin-orbit splittings that exist between spin-up and spin-down conduction bands ($\Delta_c \sim 30$~meV) and valence bands ($\Delta_v \sim 400$~meV) at the $K/K'$ points in monolayer WSe$_2$.  $\Delta_c$ and $\Delta_v$ impose huge  out-of-plane effective magnetic fields in the range of $100-1000$~T that are `seen' by electrons and holes residing at the $K$ and $K'$ points of the Brillouin zone. As such, the addition of small $B_\perp \approx 10$~mT should be insignificant in comparison. The experimental fact that small $B_\perp$ \textit{does} strongly influence $\tau_v$ (at the lowest carrier densities in sample A, and over a wider density range in sample B) therefore suggests that some population of intermediate in-gap states likely participates in the $K \leftrightarrow K'$ valley relaxation process -- and moreover, that these states are \textit{not} subject to the huge effective spin-orbit magnetic fields in monolayer TMDs, possibly because they lack the inversion-asymmetric crystal symmetry of the host lattice \cite{Kaasbjerg:2017, Abramson:2018}. Such states would have nearly degenerate spin levels.  A potential valley relaxation pathway can therefore arise if a spin-polarized carrier from (say) the $K$ valley scatters to an in-gap state and subsequently precesses about any residual in-plane effective magnetic field, and then scatters to the $K'$ valley when its spin projection has reversed.

We note that a related intermediate-state mechanism was recently proposed to account for the circular polarization of PL from low-energy defect states in monolayer WSe$_2$ \cite{Smolenski:2016}, which was found to be similarly sensitive to small $B_\perp$. As also noted in \cite{Smolenski:2016}, the sensitivity to small applied $B_\perp$ is reminiscent of early optical orientation studies of spin-polarized resident electrons in $n$-type GaAs \cite{OO, Furis:2007}, when small \textit{in-plane} effective magnetic fields were present due to Rashba or strain-related spin-orbit coupling. While such in-plane effective fields can depolarize out-of-plane oriented spins by inducing spin precession, their influence is readily suppressed by application of $B_\perp$, which stabilizes out-of-plane spin orientation.

Taken together, the measured dependence of $\tau_v$ on carrier density, encapsulation, temperature, and $B_\perp$ are consistent with a picture of electron/hole valley relaxation shown in the diagrams in Figure 2.  Once a valley polarization of the resident carriers is established by the pump pulse (following rapid exciton scattering and recombination on short timescales), the resident carrier polarization can relax either via intrinsic pathways related to density- and temperature-dependent carrier-phonon coupling, and associated Dyakonov-Perel and Elliott-Yafet mechanisms (solid lines) \cite{Song:2013, Gunst:2016, Ochoa:2014, WangWu:2014, Xu:2020}, or it can also relax via extrinsic pathways (dotted lines) mediated by a population of in-gap states that exist due to localized defects \cite{Kaasbjerg:2017, Abramson:2018} or to the disorder potential arising from an imperfect underlying substrate. If, as the dependence on $B_\perp$ suggests, these in-gap states are not subject to the huge perpendicular effective magnetic fields from $\Delta_{c,v}$ \cite{Kaasbjerg:2017, Abramson:2018}, then spin degrees of freedom at these localized sites are nearly degenerate and spins can depolarize due to precession about any residual effective fields, thereby mediating alternative relaxation pathways between the spin-valley locked carriers at $K$ and $K'$, in a two-step process depicted by the dotted lines. Note that in early studies of unpassivated CVD-grown TMD monolayers having significant disorder, evidence for a localized sub-population of carriers exhibiting free spin precession at a frequency proportional only to an applied transverse field (\textit{i.e.}, no evidence of any large effective spin-orbit field) was reported in \cite{Yang_NL:2015} and \cite{McCormick:2017, Volmer:2017} by TRKR methods.

We argue that in lower-quality TMD monolayers such as sample B (which is in direct contact with SiO$_2$), in-gap states are likely more numerous and are more broadly distributed in energy -- and therefore play a more prominent role -- as compared with fully-encapsulated high-quality monolayers such as sample A. Furthermore, an ensemble of in-gap states can be expected to play a more dominant role at low carrier densities when their distribution is only partially occupied, and a less important role at high carrier densities when they are full. This is consistent with the data shown in Fig. 2, where the multiexponential decays observed in sample B are most clearly evident at low electron and hole densities, but at higher carrier densities when in-gap states fill, the decays become more closely monoexponential (and also faster, as density-dependent intrinsic mechanisms begin to dominate). In comparison, the fully-encapsulated sample A exhibits monoexponential decays at all carrier densities, suggesting mostly intrinsic mechanisms at play. 

Consequently, and in summary, the data therefore suggest that the marked carrier density dependence of $\tau_v$ in sample A -- which drops by over two orders of magnitude from $\sim$10~$\mu$s at low density to $\sim$100~ns when $n,p \gtrsim 2 \times 10^{12}$~cm$^{-2}$ (see Fig. 2c) -- is likely an intrinsic effect, due to a Fermi energy (and momentum) -dependent valley relaxation in monolayer TMDs. It is interesting that both electron and hole doping have such similar effects on $\tau_v$; however, at the carrier densities and low temperatures (5.8~K) at which this study was performed, both the Fermi energy ($\lesssim$15~meV) and the thermal energy ($k_B T \approx 0.5$~meV) are always less than the spin-orbit splitting of the conduction bands in WSe$_2$ ($\Delta_c \approx 30$~meV). The surprising dependence of $\tau_v$ on small out-of-plane magnetic fields $B_\perp$ provides additional key insight, both by revealing the likely presence of intermediate in-gap states having negligible (or small in-plane) spin-orbit effective fields that can mix spins via precession and thereby mediate $K \leftrightarrow K'$ intervalley relaxation, and also by providing a simple means to suppress their influence so that intrinsic mechanisms can be more clearly revealed, as demonstrated in Fig. 5. Finally, the important role of substrate (dielectric) disorder on the optical properties and the valley relaxation of resident carriers is more clearly revealed by a direct side-by-side comparison of fully- and partially-encapsulated monolayers.

Looking forward, the steady improvement of TMD monolayer material quality and passivation methods can be expected to result in correspondingly enhanced valley lifetimes for resident carriers.  The nearly-monoexponential decays observed in sample A, along with the relatively small influence of out-of-plane applied magnetic fields, suggests that intrinsically-limited valley lifetimes are already being approached in fully hBN encapsulated monolayers. An improved theoretical understanding of  intrinsic valley relaxation mechanisms, and (especially) the strong temperature dependence and marked dependence on carrier density, will be extremely helpful in this regard.  Systematic experimental studies of carrier valley lifetimes in other members of the TMD monolayer family are also warranted, particularly in Mo-based monolayers where the spin-orbit splitting of the conduction band can be much smaller than in their W-based counterparts, such that even low electron density Fermi seas can occupy both spin bands in each valley. Tunable manipulation of valley degrees of freedom for resident electrons and holes remains a significant challenge, but these results suggest a potential route based on intermediate states having different symmetry, at which spin degrees of freedom can in principle be more readily controlled.

Jing Li and M. Goryca contributed equally to this work. We gratefully acknowledge Junho Choi, Cedric Robert, Bernhard Urbaszek, Xavier Marie, and Yuan Ping for helpful discussions. Work at the NHMFL was supported by the Los Alamos LDRD program. The NHMFL was supported by National Science Foundation (NSF) DMR-1644779, the State of Florida, and the U.S. Department of Energy (DOE). S.T. acknowledges support from DOE-SC0020653, Applied Materials Inc., NSF DMR-1955889, NSF CMMI-1933214, NSF 1904716, NSF 1935994, and NSF DMR-1552220.


\begin{thebibliography}{10}

\bibitem{Xu:2014} X. Xu, W. Yao, D. Xiao, and T. F. Heinz, Spin and pseudospins in layered transition metal dichalcogenides, Nat. Phys. \textbf{10}, 343 (2014).

\bibitem{Schaibley:2016} J. R. Schaibley, H. Yu, G. Clark, P. Rivera, J. S. Ross, K. L. Seyler, W. Yao, and X. Xu, Valleytronics in 2D materials, Nat. Rev. Mater. \textbf{1}, 16055 (2016).

\bibitem{Mak:2018}K. F. Mak, D. Xiao, and J. Shan, Light-valley interactions in 2D semiconductors. Nat. Photon. \textbf{12}, 451 (2018).

\bibitem{Shkolnikov:2002} Y.~P. Shkolnikov, E.~P. De~Poortere, E.~Tutuc, M.~Shayegan, Valley splitting of AlAs two-dimensional electrons in a perpendicular magnetic field, Phys. Rev. Lett. \textbf{89}, 226805 (2002).

\bibitem{Takashina:2006} K. Takashina, Y. Ono, A. Fujiwara, Y. Takahashi, and Y. Hirayama, Valley Polarization in Si(100) at Zero Magnetic Field, Phys. Rev. Lett.  \textbf{96}, 236801 (2006).

\bibitem{Rycerz:2007} A.~Rycerz, J.~Tworzyd\l{}o, C.~W.~J. Beenakker, Valley filter and valley valve in graphene, Nat. Phys. \textbf{3}, 172 (2007).

\bibitem{Xiao:2007} D. Xiao, W. Yao, and Q. Niu, Valley-contrasting physics in graphene: Magnetic moment and topological transport, Phys. Rev. Lett. \textbf{99}, 236809 (2007).

\bibitem{Yao:2008}W. Yao, D. Xiao, Q. Niu, Valley-dependent optoelectronics from inversion symmetry breaking, Phys. Rev. B \textbf{77}, 235406 (2008).

\bibitem{Xiao:2012} D. Xiao, G.-B. Liu, W. Feng, X. Xu, and W. Yao, Coupled spin and valley physics in monolayers of MoS$_2$ and other group-VI dichalcogenides, Phys. Rev. Lett. \textbf{108}, 196802 (2012).

\bibitem{OO}\textit{Optical Orientation}, F. Meier and B. P. Zakharchenya, eds. (North-Holland, Amsterdam, 1984).

\bibitem{SS}\textit{Semiconductor Spintronics and Quantum Computation}, D. D. Awschalom, D. Loss, N. Samarth, eds. (Springer-Verlag, Berlin, 2002).

\bibitem{Dyakonovbook}\textit{Spin Physics in Semiconductors}, M. I. Dyakonov, ed. (Springer-Verlag, Berlin, 2008).

\bibitem{Zeng:2012} H. Zeng, J. Dai, W. Yao, D. Xiao, and X. Cui, Valley polarization in MoS$_2$ monolayers by optical pumping, Nat. Nanotech. \textbf{7}, 490  (2012).

\bibitem{Cao:2012} T. Cao, G. Wang, W. Han, H. Ye, C. Zhu, J. Shi, Q. Niu, P. Tan, E. Wang, B. Liu, and J. Feng, Valley-selective circular dichroism of monolayer molybdenum disulphide, Nat. Commun. \textbf{3}, 887 (2012).

\bibitem{Mak:2012}  K. F. Mak, K. He, J. Shan, and T. F. Heinz, Control of valley polarization in monolayer MoS$_2$ by optical helicity, Nat. Nanotechnol. \textbf{7}, 494  (2012).

\bibitem{Sallen:2012} G. Sallen, L. Bouet, X. Marie, G. Wang, C. R. Zhu, W. P. Han, Y. Lu, P. H. Tan, T. Amand, B. L. Liu, and B. Urbaszek, Robust optical emission polarization in MoS$_2$ monolayers through selective valley excitation, Phys. Rev. B \textbf{86}, 081301(R) (2012).

\bibitem{Robert:2016} C. Robert, D. Lagarde, F. Cadiz, G. Wang, B. Lassagne, T. Amand, A. Balocchi, P. Renucci, S. Tongay, B. Urbaszek, and X. Marie, Exciton radiative lifetime in transition metal dichalcogenide monolayers, Phys. Rev. B \textbf{93}, 205423 (2016). 

\bibitem{Wang:2014} G. Wang, L. Bouet, D. Lagarde, M. Vidal, A. Balocchi, T. Amand, X. Marie, and B. Urbaszek, Valley dynamics probed through charged and neutral exciton emission in monolayer WSe$_2$, Phys. Rev. B \textbf{90}, 075413 (2014).

\bibitem{Wang:2018} G. Wang, A. Chernikov, M. M. Glazov, T. F. Heinz, X. Marie, T. Amand, and B. Urbaszek, Excitons in atomically thin transition metal dichalcogenides, Rev. Mod. Phys. \textbf{90}, 021001 (2018). 

\bibitem{Mai:2014} C. Mai, A. Barrette, Y. Yu, Y. G. Semenov, K. W. Kim, L. Cao, and K. Gundogdu, Many-body effects in valleytronics: Direct measurement of valley lifetimes in single-layer MoS$_2$, Nano Lett. \textbf{14}, 202 (2014).

\bibitem{Zhu:2014} C. R. Zhu, K. Zhang, M. Glazov, B. Urbaszek, T. Amand, Z. W. Ji, B. L. Liu, and X. Marie, Exciton valley dynamics probed by Kerr rotation in WSe$_2$ monolayers, Phys. Rev. B \textbf{90}, 161302 (2014).

\bibitem{Yu:2014} T. Yu and M. W. Wu, Valley depolarization due to intervalley and intravalley electron-hole exchange interactions in monolayer MoS$_2$, Phys. Rev. B \textbf{89}, 205303 (2014).

\bibitem{Singh:2016} A. Singh, K. Tran, M. Kolarczik, J. Seifert, Y. Wang, K. Hao, D. Pleskot, N. M. Gabor, S. Helmrich, N. Owschimikow, U. Woggon, and X. Li, Long-Lived Valley Polarization of Intravalley Trions in Monolayer WSe$_2$, Phys. Rev. Lett. \textbf{117}, 257402 (2016).

\bibitem{Hao:2016} K. Hao, G. Moody, F. Wu, C. K. Dass, L. Xu, C.-H. Chen, L. Sun, M.-Y. Li, L.-J. Li, A. H. MacDonald, and X. Li, Direct measurement of exciton valley coherence in monolayer WSe$_2$, Nat. Phys. \textbf{12}, 677 (2016).

\bibitem{DalConte:2015} S. Dal Conte, F. Bottegoni, E. A. A. Pogna, D. De Fazio, S. Ambrogio, I. Bargigia, C. D'Andrea, A. Lombardo, M. Bruna, F. Ciccacci, A. C. Ferrari, G. Cerullo, and M. Finazzi, Ultrafast valley relaxation dynamics in monolayer MoS$_2$ probed by nonequilibrium optical techniques, Phys. Rev. B \textbf{92}, 235425 (2015).

\bibitem{Yan:2017} T. Yan, S. Yang, D. Li, and X. Cui, Long valley relaxation time of free carriers in monolayer WSe$_2$, Phys. Rev. B \textbf{95}, 241406 (2017). 

\bibitem{Plechinger:2016} G. Plechinger, P. Nagler, A. Arora, R. Schmidt, A. Chernikov, A. Granados del Aguila, P. C. M. Christianen, R. Bratschitsch, C. Sch\"{u}ller, and T. Korn, Trion fine structure and coupled spin-valley dynamics in monolayer tungsten disulfide, Nat. Commun. \textbf{7}, 12715 (2016). 

\bibitem{Molina:2017} A. Molina-S\'{a}nchez, D. Sangalli, L. Wirtz, and A. Marini, Ab initio calculations of ultrashort carrier dynamics in two-dimensional materials: Valley depolarization in single-layer WSe$_2$, Nano Lett. \textbf{17}, 4549 (2017). 

\bibitem{Kroutvar:2004}M. Kroutvar, Y. Ducommun, D. Heiss, M. Bichler, D. Schuh, G. Abstreiter, J. J. Finley, Optically programmable electron spin memory using semiconductor quantum dots, Nature \textbf{432}, 81 (2004).

\bibitem{Lou:2007} X. Lou, C. Adelmann, S. A. Crooker, E. S. Garlid, J. Zhang, K. S. M. Reddy, S. D. Flexner, C. J. Palmstr\o{}m, and P. A. Crowell, Electrical Detection of Spin Transport in Lateral Ferromagnet-Semiconductor Devices, Nat. Phys. \textbf{3}, 197 (2007).

\bibitem{Appelbaum:2007} I. Appelbaum, B. Huang, and D. J. Monsma, Electronic measurement and control of spin transport in silicon, Nature \textbf{447}, 295 (2007).

\bibitem{Han:2014} W. Han, R. K. Kawakami, M. Gmitra, and J. Fabian, Graphene spintronics, Nat. Nanotech. \textbf{9}, 794 (2014).

\bibitem{Crooker:1997} S. A. Crooker, D. D. Awschalom, J. J. Baumberg, F. Flack, and N. Samarth, Optical spin resonance and transverse spin relaxation in magnetic semiconductor quantum wells, Phys. Rev. B \textbf{56}, 7574 (1997). 

\bibitem{Greilich:2006} A. Greilich, D. R. Yakovlev, A. Shabaev, Al. L. Efros, I. A. Yugova, R. Oulton, V. Stavarache, D. Reuter, A. Wieck, M. Bayer, Mode Locking of Electron Spin Coherences in Singly Charged Quantum Dots, Science \textbf{313}, 341 (2006).

\bibitem{Yang:2015} L. Yang, N. A. Sinitsyn, W. Chen, J. Yuan, J. Zhang, J. Lou, and S. A. Crooker, Long-lived nanosecond spin relaxation and spin coherence of electrons in monolayer MoS$_2$ and WS$_2$, Nat. Phys. \textbf{11}, 830 (2015).

\bibitem{Hsu:2015} W.-T. Hsu, Y.-L. Chen, C.-H. Chen, P.-S. Liu, T.-H. Hou, L.-J. Li, and W.-H. Chang, Optically initialized robust valley-polarized holes in monolayer WSe$_2$, Nat. Commun. \textbf{6}, 8963 (2015).

\bibitem{Song:2016} X. Song, S. Xie, K. Kang, J. Park, and V. Sih, Long-lived hole spin/valley polarization probed by Kerr rotation in monolayer WSe$_2$, Nano Lett. \textbf{16}, 5010 (2016).

\bibitem{McCormick:2017} E. J. McCormick, M. J. Newburger, Y. K. Luo, K. M. McCreary, S. Singh, I. B. Martin, E. J. Cichewicz, B. T. Jonker, and R. K. Kawakami, Imaging spin dynamics in monolayer WS$_2$ by time-resolved Kerr rotation microscopy, 2D Mater. \textbf{5}, 011010 (2017).

\bibitem{Robert:2017} C. Robert, T. Amand, F. Cadiz, D. Lagarde, E. Courtade, M. Manca, T. Taniguchi, K. Watanabe, B. Urbaszek, and X. Marie, Fine structure and lifetime of dark excitons in transition metal dichalcogenide monolayers, Phys. Rev. B \textbf{96}, 155423 (2017). 

\bibitem{Zhang:2017} X.-X. Zhang, T. Cao, Z. Lu, Y.-C. Lin, F. Zhang, Y. Wang, Z. Li, J. C. Hone, J. A. Robinson, D. Smirnov, S. G. Louie, and T. F. Heinz, Magnetic brightening and control of dark excitons in monolayer WSe$_2$, Nat. Nanotech. \textbf{12}, 883 (2017). 

\bibitem{Liu:2019} E. Liu, J. van Baren, Z. Lu, M. M. Altaiary, T. Taniguchi, K. Watanabe, D. Smirnov, and C. H. Lui, Gate Tunable Dark Trions in Monolayer WSe$_2$, Phys. Rev. Lett. \textbf{123}, 027401 (2019). 

\bibitem{Molas:2019} M. R. Molas, A. O. Slobodeniuk, T. Kazimierczuk, K. Nogajewski, M. Bartos, P. Kapuscinski, K. Oreszczuk, K. Watanabe, T. Taniguchi, C. Faugeras, P. Kossacki, D. M. Basko, and M. Potemski, Probing and Manipulating Valley Coherence of Dark Excitons in Monolayer WSe$_2$, Phys. Rev. Lett. \textbf{123}, 096803 (2019).

\bibitem{Chen:2020} S.-Y. Chen, M. Pieczarka, M. Wurdack, E. Estrecho, T. Taniguchi, K. Watanabe, J. Yan, E. A. Ostrovskaya, M. S. Fuhrer, Long-lived populations of momentum- and spin-indirect excitons in monolayer WSe$_2$, preprint at arXiv:2009.09602

\bibitem{Dey:2017} P. Dey, L. Yang, C. Robert, G. Wang, B. Urbaszek, X. Marie, and S. A. Crooker, Gate-controlled spin-valley locking of resident carriers in WSe$_2$ monolayers, Phys. Rev. Lett. \textbf{119}, 137401 (2017).

\bibitem{Kim:2017} J. Kim, C. Jin, B. Chen, H. Cai, T. Zhao, P. Lee, S. Kahn, K. Watanabe, T. Taniguchi, S. Tongay, M. F. Crommie, and F. Wang, Observation of ultralong valley lifetime in WSe$_2$/MoS$_2$ heterostructures, Sci. Adv. \textbf{3}, e1700518 (2017).

\bibitem{Cadiz:2017} F. Cadiz, E. Courtade, C. Robert, G. Wang, Y. Shen, H. Cai, T. Taniguchi, K. Watanabe, H. Carrere, D. Lagarde, M. Manca, T. Amand, P. Renucci, S. Tongay, X. Marie, and B. Urbaszek, Excitonic Linewidth Approaching the Homogeneous Limit in MoS$_2$-Based van der Waals Heterostructures, Phys. Rev. X \textbf{7}, 021026 (2017).

\bibitem{Ajayi:2017} O. A. Ajayi, J. V. Ardelean, G. D. Shepard, J. Wang, A. Antony, T. Taniguchi, K. Watanabe, T. F. Heinz, S. Strauf, X.-Y. Zhu, and J. C. Hone, Approaching the intrinsic photoluminescence linewidth in transition metal dichalcogenide monolayers, 2D Mater. \textbf{4}, 031011 (2017).

\bibitem{Wang:2017} Z. Wang, J. Shan, and K. F. Mak, Valley- and spin-polarized Landau levels in monolayer WSe$_2$, Nat. Nanotech. \textbf{12}, 144 (2017).

\bibitem{VanTuan:2019} D. Van Tuan, B. Scharf, Z. Wang, J. Shan, K. F. Mak, I. \v{Z}uti\'{c}, and H. Dery, Probing many-body interactions in monolayer transition-metal dichalcogenides, Phys. Rev. B \textbf{99}, 085301 (2019).

\bibitem{Jin:2018} C. Jin, J. Kim, M. I. B. Utama, E. C. Regan, H. Kleemann, H. Cai, Y. Shen, M. J. Shinner, A. Sengupta, K. Watanabe, T. Taniguchi, S. Tongay, A. Zettl, and F. Wang, Imaging of pure spin-valley diffusion current in WS$_2$-WSe$_2$ heterostructures, Science \textbf{360}, 893 (2018).

\bibitem{Goryca:2019} M. Goryca, N. P. Wilson, P. Dey, X. Xu, and S. A. Crooker, Detection of thermodynamic ``valley noise'' in monolayer semiconductors: Access to intrinsic valley relaxation time scales, Sci. Adv. \textbf{5}, eaau4899 (2019).

\bibitem{Volmer:2017} F. Volmer, S. Pissinger, M. Ersfeld, S. Kuhlen, C. Stampfer, and B. Beschoten, Intervalley dark trion states with spin lifetimes of 150 ns in WSe$_2$, Phys. Rev. B \textbf{95}, 235408 (2017).

\bibitem{Edelberg:2019} D. Edelberg, D. Rhodes, A. Kerelsky, B. Kim, J. Wang, A. Zangiabadi, C. Kim, A. Abhinandan, J. Ardelean, M. Scully, D. Scullion, L. Embon, R. Zu, E. J. G. Santos, L. Balicas, C. Marianetti, K. Barmak, X. Zhu, J. Hone, and A. N. Pasupathy, Approaching the intrinsic limit in transition metal diselenides via point defect control, Nano Lett. \textbf{19}, 4371 (2019).

\bibitem{Dean:2010} C. R. Dean, A. F. Young, I. Meric, C. Lee, L. Wang, S. Sorgenfrei, K. Watanabe, T. Taniguchi, P. Kim, K. L. Shepard, and J. Hone, Boron nitride substrates for high-quality graphene electronics, Nat. Nanotech. \textbf{5}, 722 (2010). 

\bibitem{Rhodes:2019} D. Rhodes, S. H. Chae, R. Ribeiro-Palau, and J. Hone, Disorder in van der Waals heterostructures of 2D materials, Nat. Mater. \textbf{18}, 541 (2019).

\bibitem{Raja:2019} A. Raja, L. Waldecker, J. Zipfel, Y. Cho, S. Brem, J. D. Ziegler, M. Kulig, T. Taniguchi, K. Watanabe, Ermin Malic, T. F. Heinz, T. C. Berkelbach, and A. Chernikov, Dielectric disorder in two-dimensional materials, Nat. Nanotech. \textbf{14}, 832 (2019). 

\bibitem{Martin:2020} E. W. Martin, J. Horng, H. G. Ruth, E. Paik, M. Wentzel, H. Deng, and S. T. Cundiff, Encapsulation Narrows and Preserves the Excitonic Homogeneous Linewidth of Exfoliated Monolayer MoSe$_2$, Phys. Rev. Appl. \textbf{14}, 021002 (2020). 

\bibitem{JingLi:2020} J. Li, M. Goryca, N. P. Wilson, A. V. Stier, X. Xu, and S. A. Crooker, Spontaneous Valley Polarization of Interacting Carriers in a Monolayer Semiconductor, Phys. Rev. Lett. \textbf{125}, 147602 (2020).

\bibitem{VanTuan:2017}D. Van Tuan, B. Scharf, I. \v{Z}uti\'{c}, H. Dery, Marrying Excitons and Plasmons in Monolayer Transition-Metal Dichalcogenides, Phys. Rev. X \textbf{7}, 041040 (2017).

\bibitem{Zhang:2009} Y. Zhang, V. W. Brar, C. Girit, A. Zettl, and M. F. Crommie, Origin of spatial charge inhomogeneity in graphene, Nat. Phys. \textbf{5}, 722 (2009).

\bibitem{Decker:2011} R. Decker, Y. Wang, V. W. Brar, W. Regan, H.-Z. Tsai, Q. Wu, W. Gannett, A. Zettl, and M. F. Crommie, Local electronic properties of graphene on a BN substrate via scanning tunneling microscopy, Nano Lett. \textbf{11}, 2291 (2011).

\bibitem{Shin:2016} B. G. Shin, G. H. Han, S. J. Yun, H. M. Oh, J. J. Bae, Y. J. Song, C.-Y. Park, and Y. H. Lee, Indirect bandgap puddles in monolayer MoS$_2$ by substrate-induced local strain, Adv. Mater. \textbf{28}, 9378 (2016). 

\bibitem{Borys:2017} N. J. Borys, E. S. Barnard, S. Gao, K. Yao, W. Bao, A. Buyanin, Y. Zhang, S. Tongay, C. Ko, J. Suh, A. Weber-Bargioni, J. Wu, Li Yang, and P. J. Schuck, Anomalous Above-Gap Photoexcitations and Optical Signatures of Localized Charge Puddles in Monolayer Molybdenum Disulfide, ACS Nano \textbf{11}, 2115 (2017).

\bibitem{Ersfeld:2020} M. Ersfeld, F. Volmer, L. Rathmann, L. Kotewitz, M. Heithoff, M. Lohmann, B. Yang, K. Watanabe, T. Taniguchi, L. Bartels, J. Shi, C. Stampfer, and B. Beschoten, Unveiling valley lifetimes of free charge carriers in monolayer WSe$_2$, Nano Lett. \textbf{20}, 3147 (2020).

\bibitem{Ochoa:2013} H. Ochoa, F. Guinea, and V. I. Fal'ko, Spin memory and spin-lattice relaxation in two-dimensional hexagonal crystals, Phys. Rev. B 88, 195417 (2013). 

\bibitem{Song:2013}Y. Song and H. Dery, Transport Theory of Monolayer Transition-Metal Dichalcogenides through Symmetry, Phys. Rev. Lett. \textbf{111}, 026601 (2013).

\bibitem{Gunst:2016} T. Gunst, T. Markussen, K. Stokbro, and M. Brandbyge, First-principles method for electron-phonon coupling and electron mobility: Applications to two-dimensional materials, Phys. Rev. B \textbf{93}, 035414 (2016).

\bibitem{Ochoa:2014} H. Ochoa, F. Finocchiaro, F. Guinea, V. I. Fal'ko, Spin-valley relaxation and quantum transport regimes in two-dimensional transition-metal dichalcogenides. Phys. Rev. B 90, 235429 (2014). 

\bibitem{WangWu:2014} L. Wang and M. W. Wu, Electron spin relaxation due to D'yakonov-Perel' and Elliot-Yafet mechanisms in monolayer MoS$_2$: Role of intravalley and intervalley processes, Phys. Rev. B \textbf{89}, 115302 (2014).

\bibitem{Xu:2020} J. Xu, A. Habib, S. Kumar, F. Wu, R. Sundararaman, and Y. Ping, Spin-phonon relaxation from a universal ab initio density-matrix approach, Nat. Commun. \textbf{11}, 2780 (2020).

\bibitem{Kaasbjerg:2017}K. Kaasbjerg, J. H. J. Martiny, T. Low, and A.-P. Jauho, Symmetry-forbidden intervalley scattering by atomic defects in monolayer transition-metal dichalcogenides, Phys. Rev. B \textbf{96}, 241411 (2017).

\bibitem{Abramson:2018} S. Refaely-Abramson, D. Y. Qiu, S. G. Louie, and J. B. Neaton, Defect-Induced Modification of Low-Lying Excitons and Valley Selectivity in Monolayer Transition Metal Dichalcogenides, Phys. Rev. Lett. \textbf{121}, 167402 (2018).

\bibitem{Smolenski:2016} T. Smole\'{n}ski, M. Goryca, M. Koperski, C. Faugeras, T. Kazimierczuk, A. Bogucki, K. Nogajewski, P. Kossacki, and M. Potemski, Tuning Valley Polarization in a WSe$_2$ Monolayer with a Tiny Magnetic Field, Phys. Rev. X \textbf{6}, 021024 (2016). 

\bibitem{Furis:2007} M. Furis, D. L. Smith, S. Kos, E. S. Garlid, K. S. M. Reddy, C. J. Palmstrom, P. A. Crowell and S. A. Crooker, Local Hanle-effect studies of spin drift and diffusion in n:GaAs epilayers and spin-transport devices, New J. Phys. \textbf{9}, 347 (2007).

\bibitem{Yang_NL:2015} L. Yang, W. Chen, K. M. McCreary, B. T. Jonker, J. Lou, and S. A. Crooker, Spin Coherence and Dephasing of Localized Electrons in Monolayer MoS$_2$, Nano Lett. \textbf{15}, 8250 (2015). 


\end{thebibliography}
\end{document}